\documentclass[epj]{webofc}
\usepackage[utf8]{inputenc}
\usepackage[varg]{txfonts}   
\usepackage{booktabs}
\usepackage{xcolor}
\usepackage{breqn}
\definecolor{darkred}{rgb}{0.4,0.0,0.0}
\definecolor{darkgreen}{rgb}{0.0,0.4,0.0}
\definecolor{darkblue}{rgb}{0.0,0.0,0.4}
\usepackage[bookmarks,linktocpage,colorlinks,
    linkcolor = darkred,
    urlcolor  = darkblue,
    citecolor = darkgreen]{hyperref}
\usepackage{subfigure}
\wocname{EPJ Web of Conferences}
\woctitle{Lattice2017}
\begin{document}
%
\selectlanguage{english}
\title{%
Spectroscopy of Charmed and Bottom Hadrons using Lattice QCD
}
\author{%
\firstname{Sourav} \lastname{Mondal}\inst{1}
\and
\firstname{M.} \lastname{Padmanath}\inst{2} \and
\firstname{Nilmani}  \lastname{Mathur}\inst{1}\fnsep\thanks{Speaker, \email{nilmani@theory.tifr.res.in}}
}
\institute{%
Department of Theoretical Physics, Tata Institute of Fundamental Research, Mumbai.
\and
Instit\"{u}t fur Theoretische Physik, Universit\"{a}t Regensburg, \\
Universit\"{a}sstrase 31,  93053 Regensburg, Germany.
}
\abstract{%
  We present preliminary results on the light, charmed and bottom baryon
  spectra using overlap valence quarks on the background of 2+1+1 flavours HISQ
  gauge configurations of the MILC collaboration. These calculations are performed on three
  different gauge ensembles at three lattice spacings ($a \sim$ 0.12 fm, 0.09 fm
  and 0.06 fm) and for physical strange, charm and bottom quark
  masses. The SU(2) heavy baryon chiral
  perturbation theory is used to extrapolate baryon masses to the
  physical pion mass and the continuum limit extrapolations are also
  performed. Our results are consistent with the well measured charmed baryons. 
  We predict the masses of many other states which are yet to be discovered.
}
\maketitle

\section{Introduction}\label{intro}
Energy spectra of heavy hadrons as well as the spin splittings between
them play an important role in understanding the fundamental strong
interactions. With the recent experimental discovery of the doubly
charmed baryon, $\Xi_{cc}^{++}$~\cite{Aaij:2017ueg}, and various
$\Omega_c^0$ resonances~\cite{Aaij:2017nav} by the LHCb collaboration,
there has been resurgence of scientific interest in the study of
heavy baryons. Anticipating discovery of many more heavy hadrons at
ongoing experiments at LHCb, BESIII and future experiments at Belle,
first principle calculations such as using Lattice QCD are essential
to investigate the energy spectra and the properties of these heavy
hadrons in a model independent way.  Not only that these calculations
are crucial in understanding the structure and interactions of these 
hadronic excitations, but also can make precise predictions that can guide future discoveries. However, lattice QCD study of heavy hadrons is severely
affected by the large discretization errors due to relatively larger
heavy quark masses. Therefore, it is important to systematically
approach the continuum limit to reduce and quantify these
discretization uncertainties. In this talk, we present updated results
from our ongoing study of the hadron spectra. Particular emphasis is
given on the calculations of the charmed and bottom baryons. Results
are extracted at three lattice spacings and then are extrapolated to
the continuum limit.

\section{Numerical Details}\label{sec-1}
We use three $N_f$=2+1+1 flavor dynamical ensembles generated by the
MILC collaboration with sizes $24^3\times64$, $32^3\times96$ and $48^3\times144$
and at gauge couplings $\beta=6.00, 6.30$ and $6.72$ respectively.  The
details of these gauge configurations are summarized in
Ref.~\cite{Bazavov:2012xda}. We use $\Omega_{sss}$ baryon mass to calculate
lattice spacings~\cite{Basak:2012py,Basak:2013oya} and those are found
to be consistent with $0.1207(11), 0.0888(8)$ and $0.0582(5)$ fm., as
measured by the MILC collaboration using $r_1$ parameter.

For the light and charm quarks we use a unified approach adopting
overlap fermions, which does not have $\mathcal{O}(ma)$ errors and has
exact chiral symmetry at finite lattice spacing. Details of the
action, its numerical implementation, mass tuning are given in
Refs.~\cite{Basak:2012py,Basak:2013oya}. The tuned bare charm quark
masses ($am_c$) are found to be 0.528, 0.425 and 0.29 on coarser to
fine lattices respectively~\cite{Basak:2013oya}.

While we intend to treat bottom quarks with the
same formalism in future, for the current work we use a non-relativistic
formulation~\cite{Lepage:1992nrqcd}. This NRQCD Hamiltonian is
improved by including spin-independent terms through
$\mathcal{O}(v^4)$. For the coarser two ensembles, we use the values
of the improvement coefficients, $c_1$ to $c_6$, as estimated
non-perturbatively by the HPQCD
collaboration~\cite{Dowdall:2011wh} on the same ensembles. 
For the fine lattice, we use
tree level coefficients. The details of the NRQCD action and tuning is
given in Ref.~\cite{Mathur:2016hsm, Lewis:2008fu}.  The spin averaged
$1S$ bottomonium kinetic mass is utilized to tune the bottom quark
mass~\cite{Mathur:2016hsm}. The observed hyperfine splitting ($64\pm
3$ MeV) in $1S$ bottomonium is found to be in good agreement with its
experimental value ($62.3\pm 3$ MeV) .

\vspace*{-0.1in}

\section{Results}\label{results}
We have calculated hadron spectra over a wide range of pseudoscalar
meson masses from the physical pion mass to pseudoscalar meson mass at
about 5.5 GeV. On fine lattices pseudoscalar meson mass ranges are
from about 300 MeV to around 6 GeV. In future we intend to extend
pseudoscalar meson mass range towards $\eta_b$ on a hyperfine
lattice. While that calculation is underway, using non-relativistic
bottom quarks we have extended this study to calculate the energy
spectra of bottom hadrons. While a comprehensive analysis with
existing data set is also underway, here we present our preliminary
results on the ground state energy spectra for light as well as heavy
hadrons with particular emphasis on charmed and charmed-bottom
baryons.

\begin{figure}[h]
\vspace{-0.1in}
\hspace{-0.3cm}
\parbox{.5\linewidth}{
\centering
\includegraphics[width=0.55\textwidth,height=0.36\textwidth,clip=true]{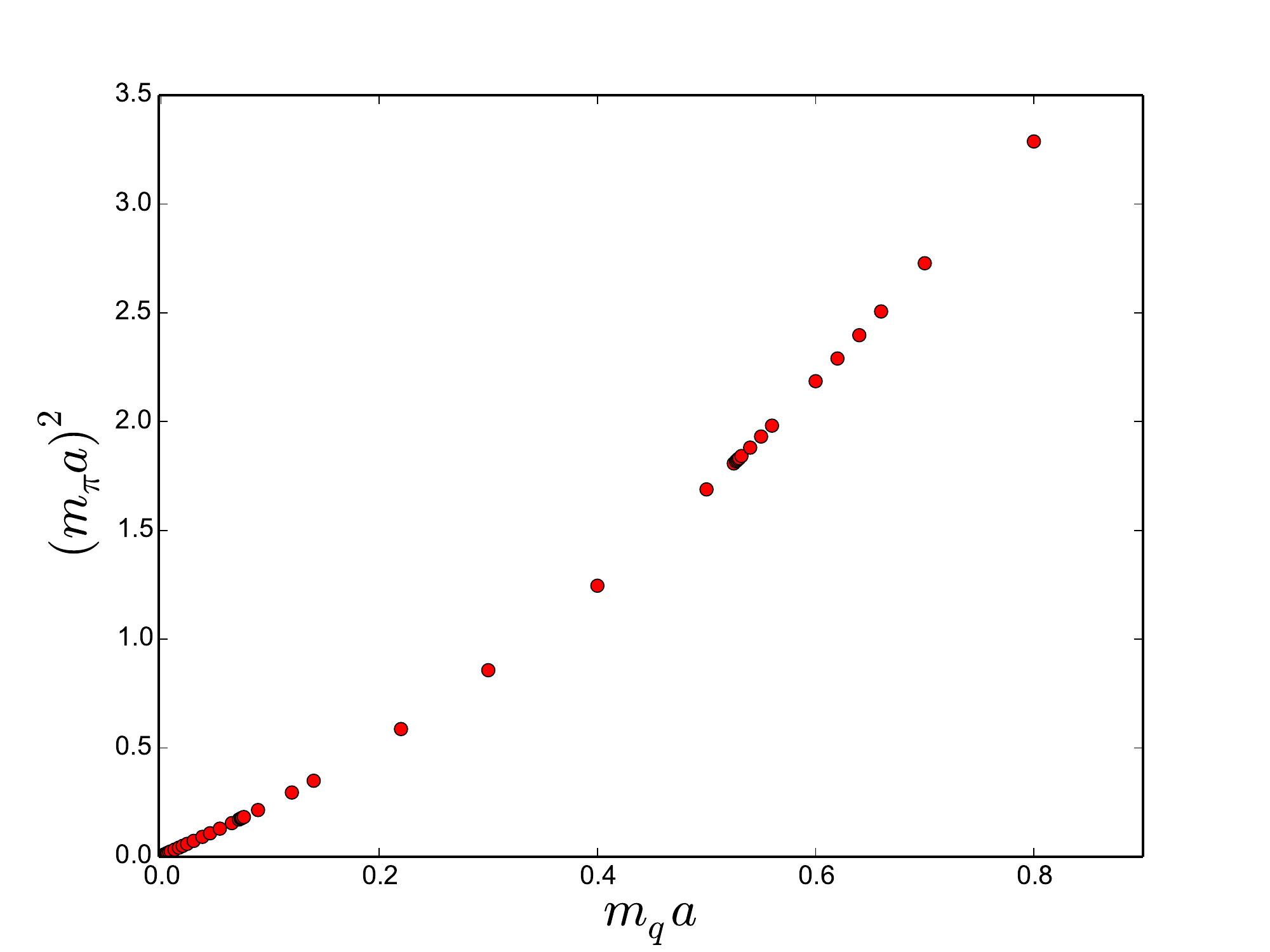}\\
(a)}
\hspace{-0.5cm}
\parbox{.5\linewidth}{
 \centering
\includegraphics[width=0.55\textwidth,height=0.36\textwidth,clip=true]{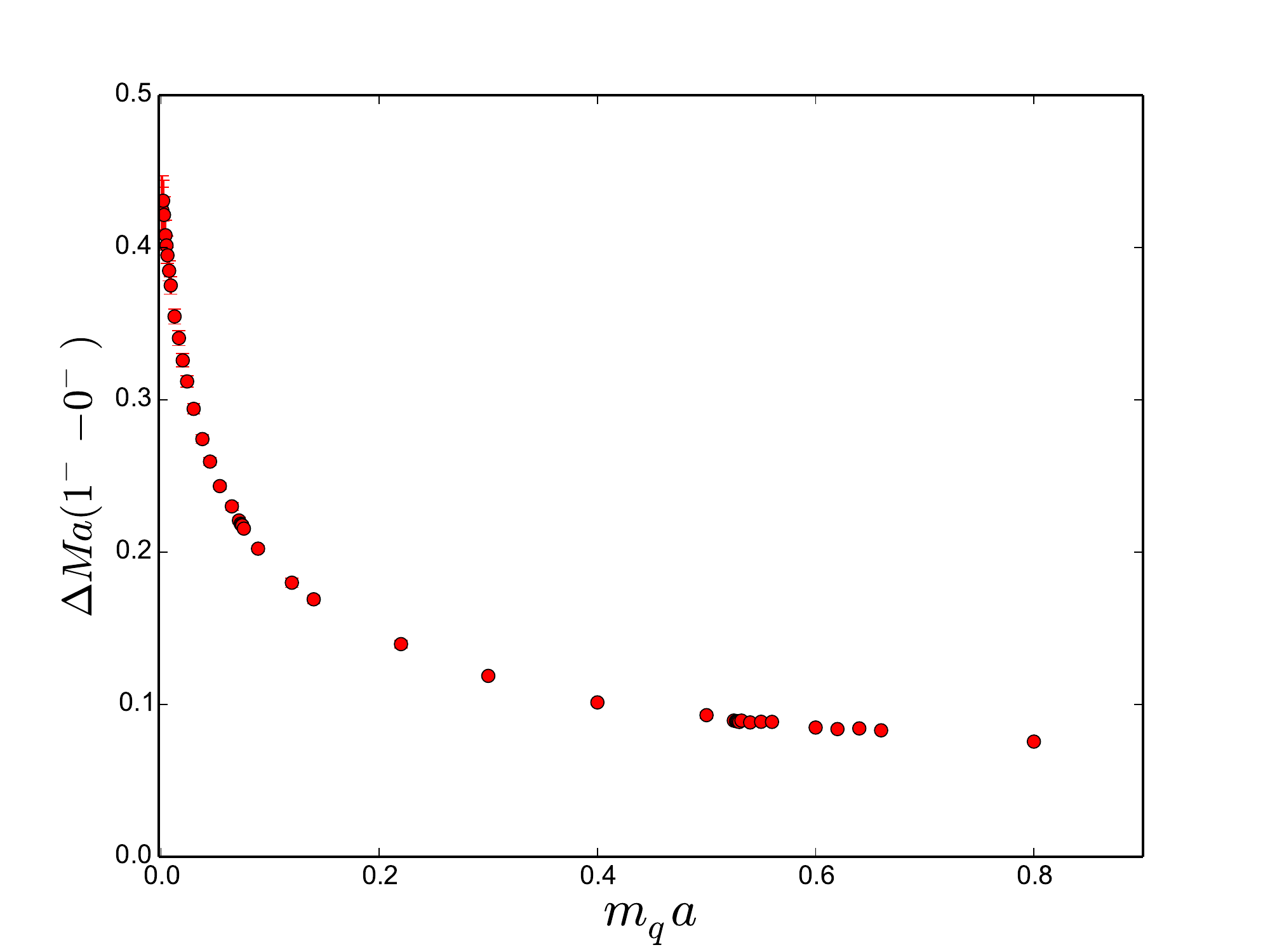}\\
(b)}
\hspace{-1.5cm}
\parbox{.5\linewidth}{
\centering
\includegraphics[width=0.55\textwidth,height=0.36\textwidth,clip=true]{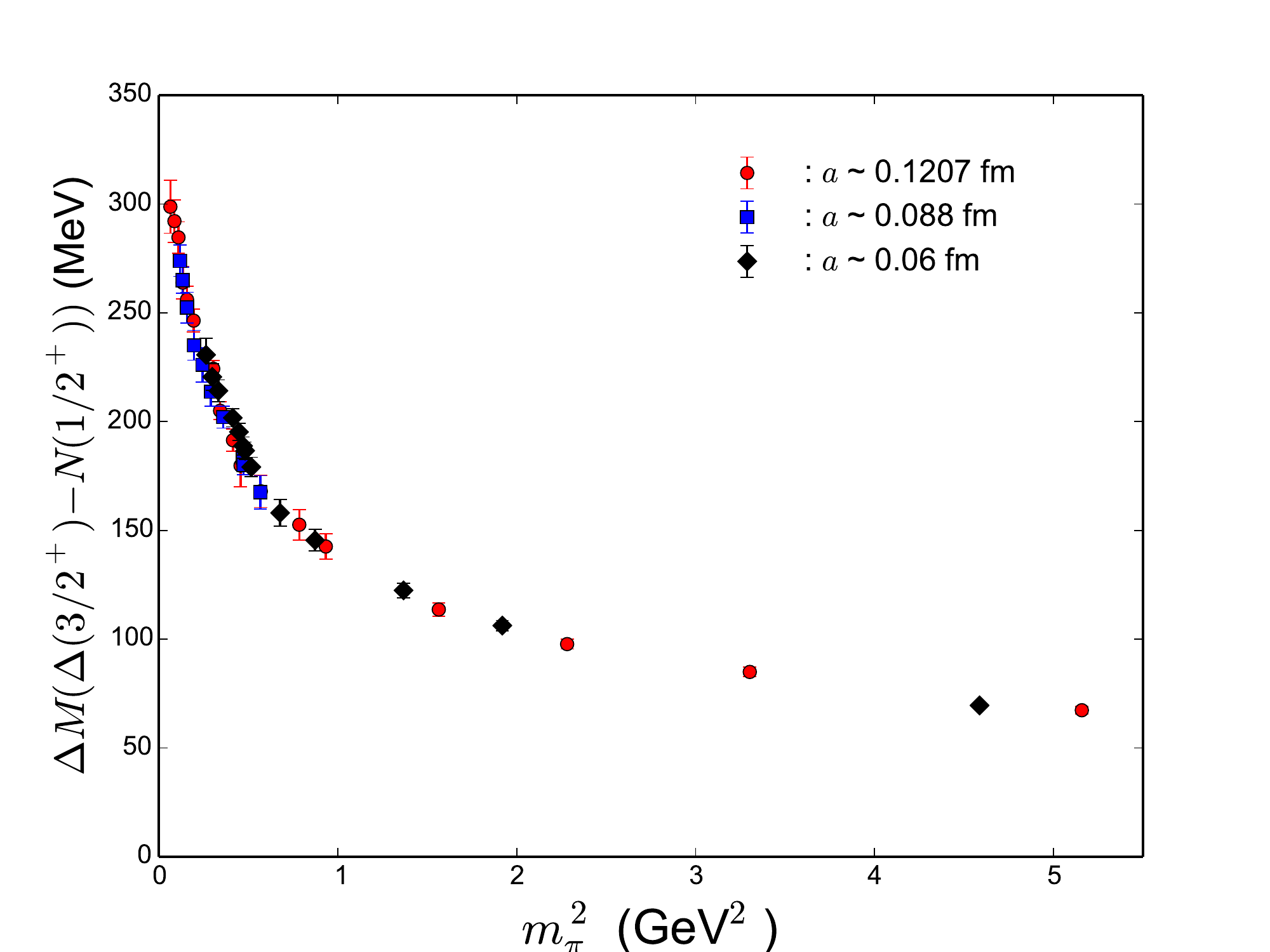}\\
(c)}
\parbox{.5\linewidth}{
 \centering
\includegraphics[width=0.55\textwidth,height=0.36\textwidth,clip=true]{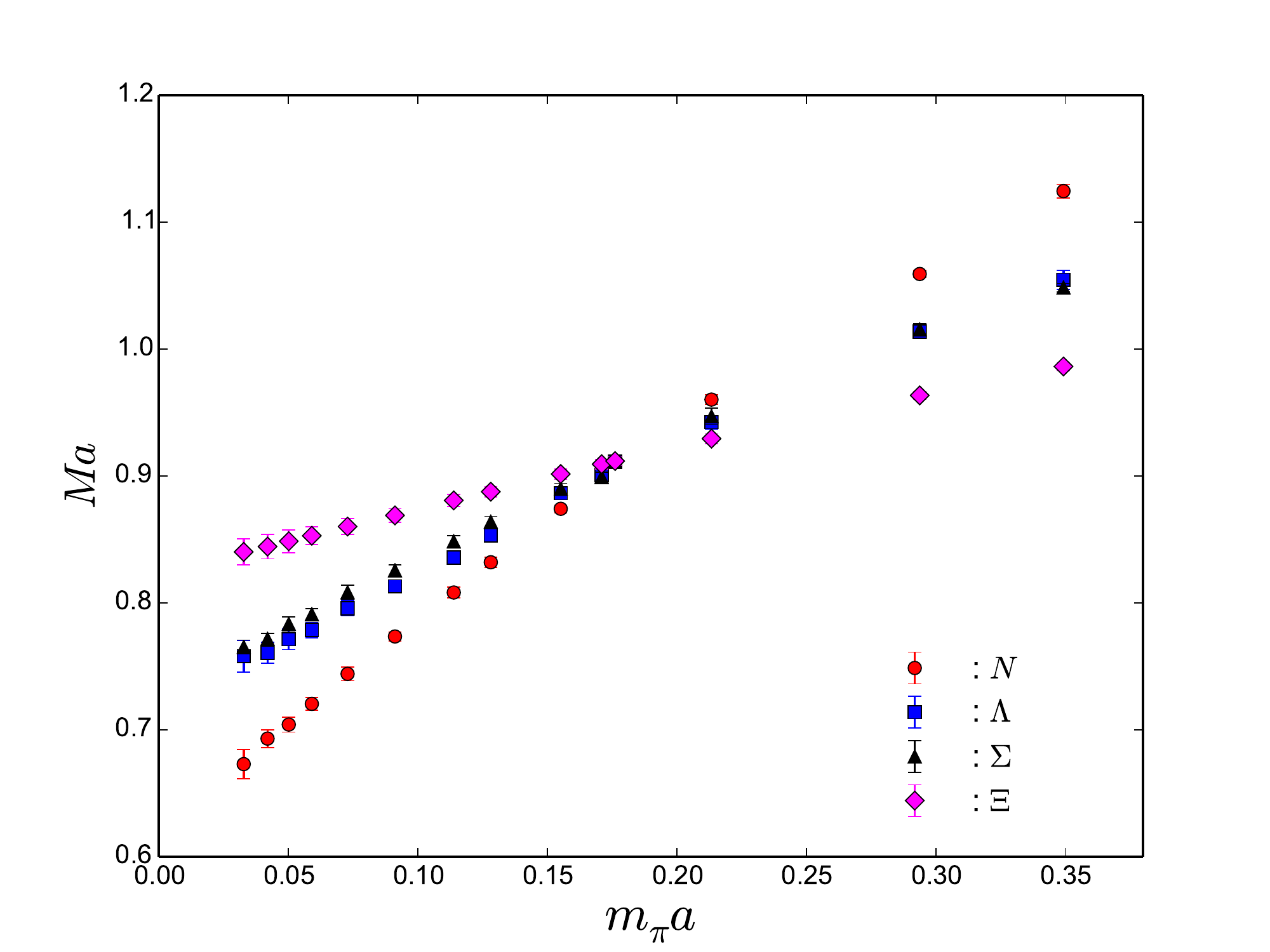}\\
(d)}
\caption{ (a) Square of pseudoscalar meson masses as a function of quark masses from physical pion mass to about 
5.5 GeV. (b) Hyperfine splittings between vector and pseudoscalar mesons over the same range of pion masses. 
(c) Hyperfine splittings between $\Delta$-baryon ($J^{P} = {3\over 2}^{+}$)  and the nucleon ($J^{P} = {1\over 2}^{+}$) 
at three lattice spacings and over a wide range of pseudoscalar meson masses. (d) Ground state energy spectra of 
the low lying octet baryons on $32^3\times 96$ ensemble. 
\label{fig_light_hadrons}}
\end{figure}
\vspace*{-0.1in}
\subsection{Light hadrons}
In Figure~\ref{fig_light_hadrons}, we present our results on 
the light hadron spectra. Figure 1(a) shows pseudoscalar meson masses as a function of quark masses on the coarser lattice ($a \sim
0.1207$ fm) covering a mass range from the physical pion mass to around 5.5 GeV. It is quite encouraging to see that the square of 
the pseudoscalar meson mass approaches the chiral limit linearly with quark mass (modulo small volume effect). In Figure 1(b) we show the hyperfine splittings between vector ($1^{-}$) and pseudoscalar ($0^{-}$) mesons on this wide range of 
pseudoscalar meson masses. In Figure 1(c) we show the similar hyperfine splittings (in MeV) between $\Delta$-baryon ($J^{P} = {3\over 2}^{+}$) 
and nucleon ($J^{P} = {1\over 2}^{+}$) at three lattice spacings and also over a wide range of pseudoscalar meson masses. Figure 1(d) 
shows our preliminary results on the ground state spectra of the low lying octet baryons on $32^3 \times 96$ lattice at the lattice spacing 
$\sim 0.088$ fm. As expected energies of these baryons coincide at the SU(3) flavour symmetric point and deviate from each 
other on its both sides. An analysis with the chiral and continuum extrapolation of these baryons is ongoing.

\subsection{Charmed hadrons}
Plethora of experimental discoveries have been made over the last two decades in the heavy hadron sector, part of which are understood 
theoretically, while the nature of the rest continues to be puzzling~\cite{Patrignani:2016xqp}. Investigations using first principle calculations, 
such as lattice QCD, are crucial to understand the structure and interactions of these excitations as well as to guide future discoveries
of more subatomic particles. As mentioned earlier, being heavy, the energy spectra of heavy hadrons on the lattice are subject to strong discretization errors 
and thus lattice calculations at more than one lattice spacings followed by a systematic continuum extrapolation is 
quite essential. 

\begin{figure}[h]
\vspace{-0.1in}
\hspace{-0.5cm}
\parbox{.5\linewidth}{
\centering
\includegraphics[width=0.56\textwidth,height=0.45\textwidth,clip=true]{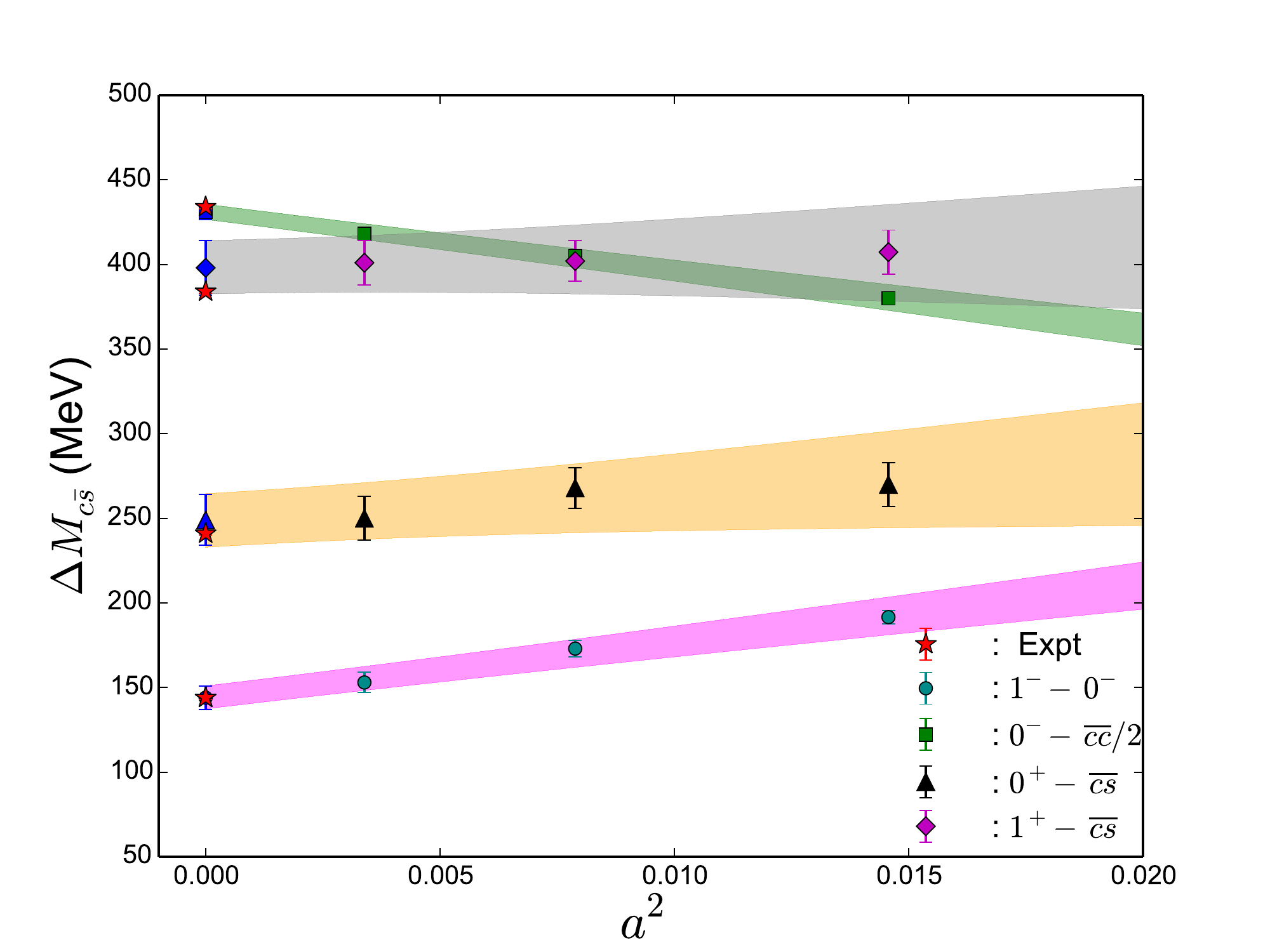}\\
(a)}
\hspace{-0.0cm}
\parbox{.5\linewidth}{
 \centering
\includegraphics[width=0.56\textwidth,height=0.45\textwidth,clip=true]{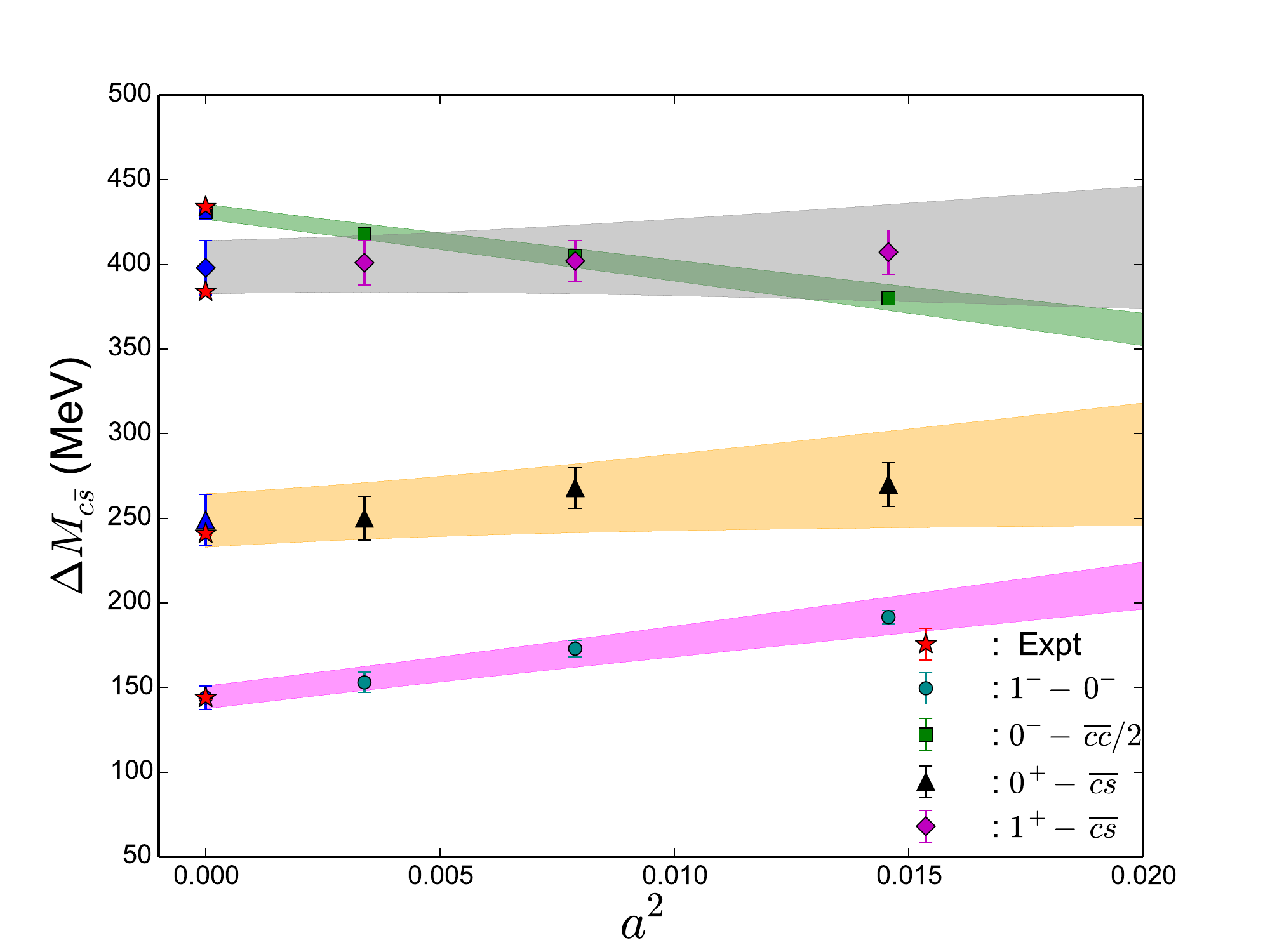}\\
(b)}
\caption{ Energy splittings in (a) the charm-strange and (b) the charmonium ground state 
spectra plotted against the square of the lattice spacings at three lattice spacings. 
Bands represent one-sigma errorbars in the continuum extrapolations. \label{fig_hyp_cs_cc}}
\end{figure}

In Figure~\ref{fig_hyp_cs_cc} we plot our preliminary results on continuum extrapolations for charm-strange mesons (Figure~\ref{fig_hyp_cs_cc}(a)) 
and charmonia (Figure~\ref{fig_hyp_cs_cc}(b)). To reduce discretization errors, we  calculate splitting of a meson 
from the respective $1S$ spin average masses. It is interesting to note that different mesons have different slopes towards the
continuum limit. The continuum extrapolation is carried out using terms up to $\mathcal O((m_qa)^3)$ with Bayesian priors (there is no 
$\mathcal O(m_qa)$ term for overlap action). We observe that except for the hyperfine splittings, the 
coefficients for $\mathcal O((m_qa)^3)$ terms are very small.

Next we present our preliminary results for 
the ground state of the positive parity charmed baryons. In order to reduce discretization errors, we perform chiral extrapolation 
on the ratios of baryon masses ($M_ba$) to the $1S$ spin-average mass ($M_{1S}a$), i.e., $M_b^r = M_ba/nM_{1S}a$, where $n = 1/2$ and 
1 for singly-and doubly charmed baryons, respectively. The chiral extrapolations are made with a naive quadratic fit form in the quark mass,
 $M_b^r = A+B.(m_{\pi}a)^2$, as well as with a chiral extrapolation form (equations below) using heavy baryon chiral perturbation theory (HBChPT), 
as described in Ref.~\cite{Briceno:2012wt}.
\begin{equation}
\frac{m_{\Lambda_c}}{m_{spin av.}}=\frac{m_{\Lambda_c}^0}{m_{spin av.}^0}+\frac{\sigma_{\Lambda_c}}{(4\pi f_{\pi})m_{spin av.}}m_{\pi}^2 - \frac{6g_3^2}{(4\pi f_{\pi})^2(m_{spin av.})}\left(\frac{1}{3}\mathcal{F}(m_{\pi},\Delta_{\Lambda_c\Sigma_c,\mu})+\frac{2}{3}\mathcal{F}(m_{\pi},\Delta_{\Lambda_c\Sigma_c^*,\mu})\right)	
\end{equation}
\begin{equation}
\frac{m_{\Xi_c}}{m_{spin av.}}=\frac{m_{\Xi_c}^0}{m_{spin av.}^0}
+\frac{\sigma_{\Xi_c}}{(4\pi f_{\pi})m_{spin av.}}m_{\pi}^2 - \frac{3}{2}\frac{g_3^2}{(4\pi f_{\pi})^2(m_{spin av.})}\left(\frac{1}{3}\mathcal{F}(m_{\pi},\Delta_{\Xi_c\Xi_c^{\prime},\mu})+\frac{2}{3}\mathcal{F}(m_{\pi},\Delta_{\Xi_c\Xi_c^*,\mu})\right)
\label{chextrplnfrm}
\end{equation}
for $\Lambda_c$ and $\Xi_c$ respectively. The chiral function $\mathcal{F}$ in eqn.~\ref{chextrplnfrm} is defined as,
\begin{equation}
\mathcal{F}(m,\Delta, \mu)=(\Delta^2-m^2+i\epsilon)^{3/2}\ln\left(\frac{\Delta + \sqrt{\Delta^2-m^2+i\epsilon}}{\Delta -\sqrt{\Delta^2-m^2+i\epsilon}}\right)-\frac{3}{2}\Delta m^2\ln\left(\frac{m^2}{\mu^2}\right)-\Delta^3\ln\left(\frac{4\Delta^2}{m^2}\right),
\end{equation}
with $\mathcal{F}(m,0,\mu)=\pi m_{\pi}^3$.
Splittings $\Delta$ used in the extrapolation formula are obtained by extrapolating the splittings between two baryons to the physical 
pion masses using trivial extrapolation form 
 \begin{equation}
 \Delta_{ij}=\Delta_{ij}^0+A \,(m_{\pi}a)^2,
 \end{equation}
where $i$ and $j$ are the baryons under consideration. For $\Lambda_c$ and $\Xi_c$ we could use HBChPT with $\chi^2/dof \sim 1$. 
We then perform continuum extrapolation of the chirally extrapolated ratios with a form up to 
$\mathcal{O}(a^2)$ terms. Finally to obtain the physical values we multiply the extrapolated values by $nM_{phy}(1S:c\overline{c})$. 

\begin{figure}[h]
\vspace{-0.1in}
\parbox{1\linewidth}{
\centering
\includegraphics[width=0.75\textwidth,height=0.5\textwidth,clip=true]{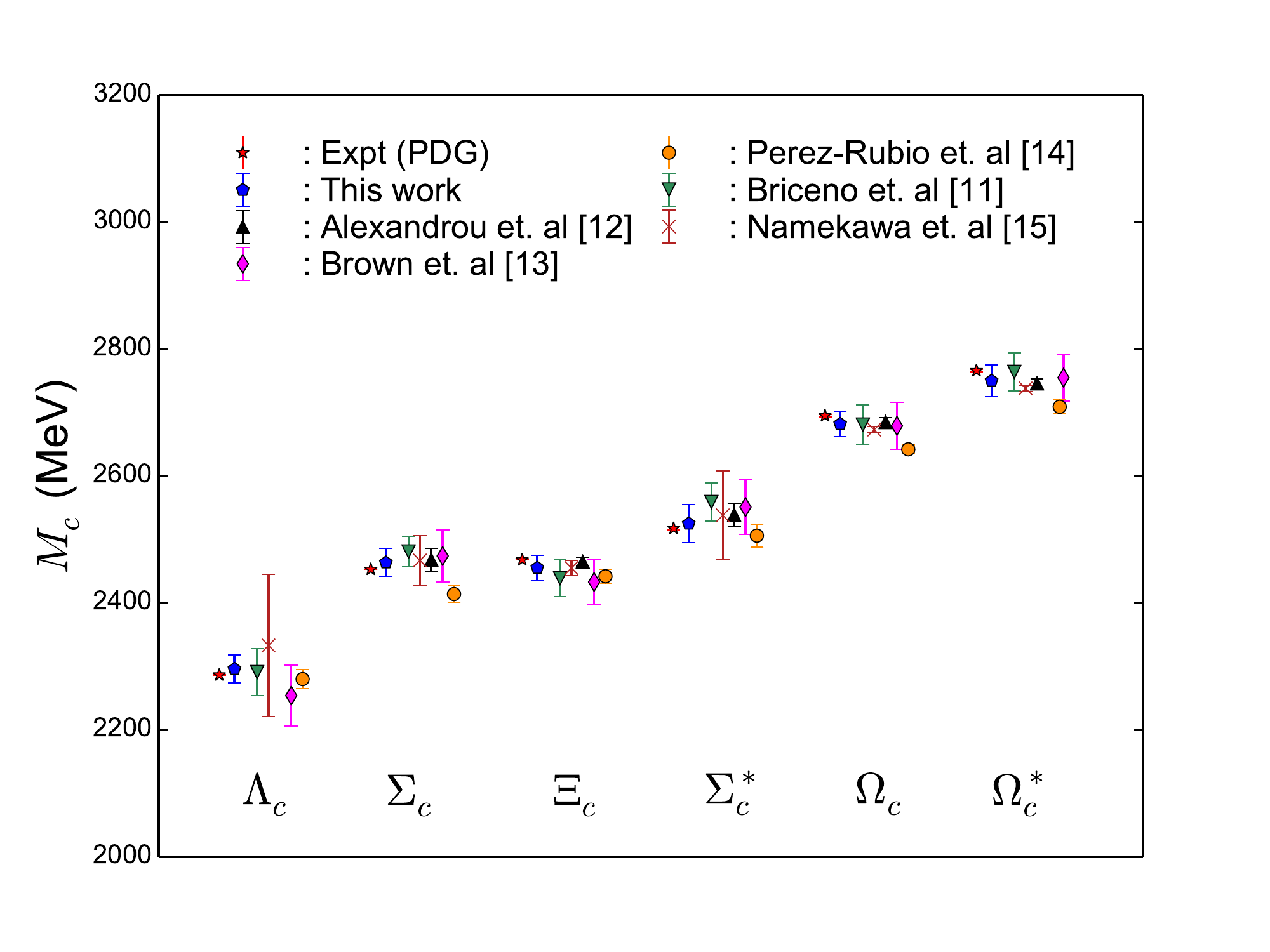}}
\vspace{-0.1in}
\caption{Preliminary results for the positive parity singly charmed baryons. Results from this work are compared with experimental values~\cite{Patrignani:2016xqp} and other lattice results~\cite{Alexandrou:2017xwd, Brown:2014ena, Bali:2015lka, Briceno:2012wt, Namekawa:2013vu}.
\label{fig_cbar}}
\end{figure}

In Figure~\ref{fig_cbar} we show our preliminary chiral and continuum extrapolated results for the ground state singly charm baryons. 
We compare our results with experimental values of these baryons~\cite{Patrignani:2016xqp} and also with other lattice results~\cite{Alexandrou:2017xwd, 
Brown:2014ena, Bali:2015lka, Briceno:2012wt, Namekawa:2013vu}. 

Here we would like to point out that the LHCb Collaboration has recently reported observation of five new resonances based on the invariant mass 
distribution of $\Xi_c^{+}K^{-}$  in  the energy  range  between  3000$-$3120  MeV~\cite{Aaij:2017nav}. These resonances have been interpreted as 
the excited states of $\Omega^{0}_{c}$ baryon. Before the discovery of these resonances we studied the excited state spectra of $\Omega^{0}_{c}$ 
baryons in detail~\cite{Padmanath:2013bla, Padmanath:2014bxa, Padmanath:2015bra}.  It is quite satisfying to see that our prediction matches 
very well with the experimental results and strongly indicates that the observed states $\Omega_c(3000)^0$ and $\Omega_c(3050)^0$ have spin-parity 
$J^P = 1/2^{-}$, the states $\Omega_c(3066)^0$ and $\Omega_c(3090)^0$ have $J^P = 3/2^{-}$, whereas $\Omega_c(3119)^0$ is possibly a $5/2^{-}$ 
state~\cite{Padmanath:2017lng}. This identification is crucial to decipher the nature of these resonances.
\begin{figure}[h]
\vspace{-0.1in}
\parbox{1\linewidth}{
\centering
\includegraphics[width=0.75\textwidth,height=0.5\textwidth,clip=true]{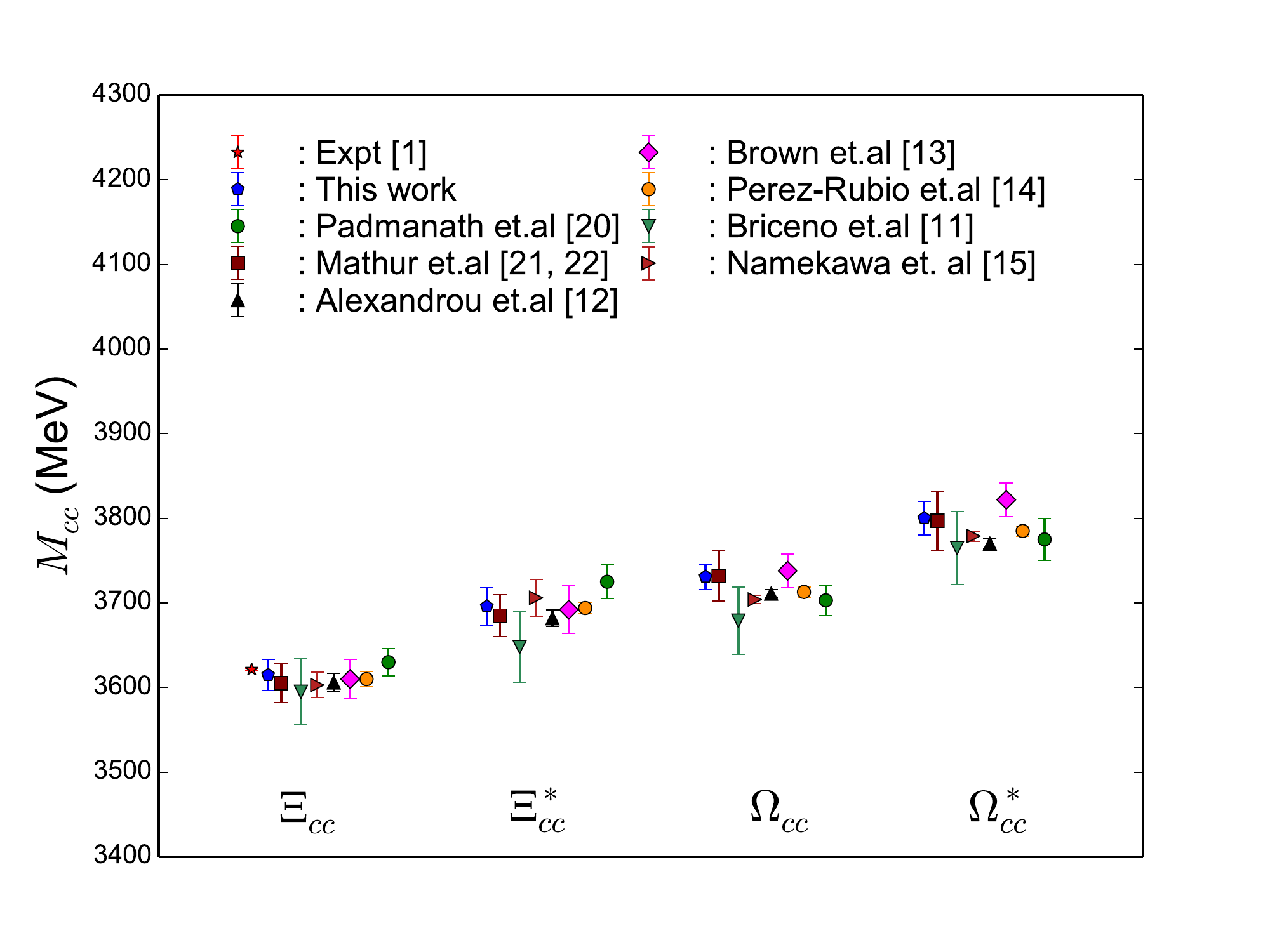}}
\vspace{-0.1in}
\caption{Preliminary results for the positive parity doubly charmed baryons. Results from this work are compared with experimental values (where available) and other lattice results~\cite{Alexandrou:2017xwd, Brown:2014ena, Bali:2015lka, Briceno:2012wt, Namekawa:2013vu, Padmanath:2015jea, Mathur:2002ce, Lewis:2001iz}
\label{fig_ccbar}}
\end{figure}

In Figure~\ref{fig_ccbar}, we show our preliminary results for doubly charmed baryons. It is noteworthy to point out that all lattice results 
including the current work are predictions before the experimental discovery of $\Xi_{cc}^{++}$ by the LHCb collaboration~\cite{Aaij:2017ueg}. 
So far $\Xi_{cc}^{++}$ is the only doubly charmed baryon discovered experimentally. In that context lattice predictions for other doubly charmed 
baryons are very interesting for their future discovery.\vspace*{-0.1in}
	
\subsection{Charmed Bottom Baryons}
Except the pseudoscalar $B_c$ and $B_c(2S)$ mesons, no other hadron has been discovered yet with $b$ and $c$ 
quark content together. Anticipating discoveries of such hadrons in the near future, we present our results on charmed-bottom hadrons in this section. 
For the calculation of $bc$ hadrons we use NRQCD propagators for bottom quarks which are contracted with overlap propagators to obtain correlators 
for various $bc$ hadrons. We have already presented our predictions for the hyperfine splitting of $B_c$ meson to be $55\pm 4$ MeV~\cite{Mathur:2016hsm}, 
which constrains the mass of the vector $B_c^*$ meson. Here we present the preliminary results on the ground state spectrum of the positive parity 
charmed bottom $\Omega$ baryons, namely, $\Omega_{ccb}({\frac{1}{2}}^+)$, $\Omega_{ccb}^*({\frac{3}{2}}^+)$, $\Omega_{cbb}({\frac{1}{2}}^+)$ and 
$\Omega_{cbb}^*({\frac{3}{2}}^+)$. Works on the spectra of $\Xi_{cbu}$ and $\Xi_{cbs}$ baryons as well on negative parity baryons are ongoing.
 
\begin{figure}[tp]
	\centering
	\subfigure[ $\Omega_{ccb}$]{\includegraphics[width=0.45\textwidth,clip]{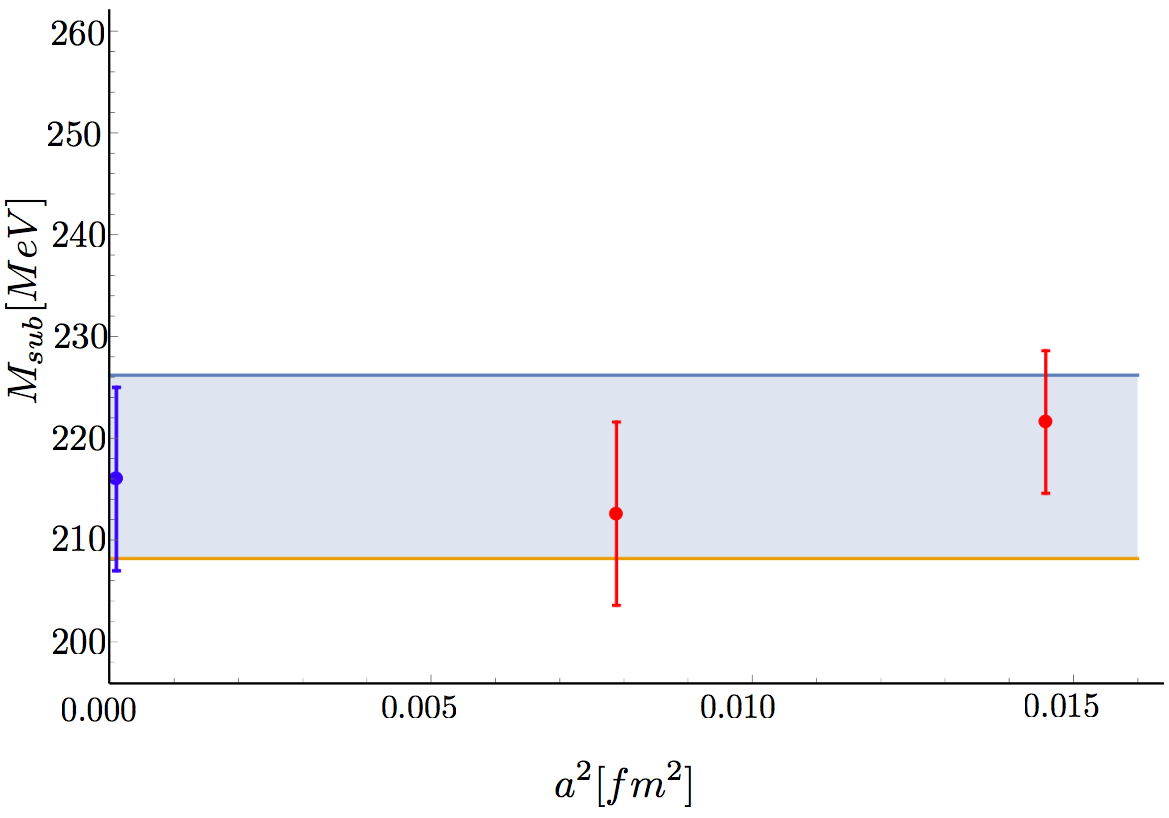}}\hfill
	\subfigure[ $\Omega_{ccb}^{*}$]{\includegraphics[width=0.45\textwidth,clip]{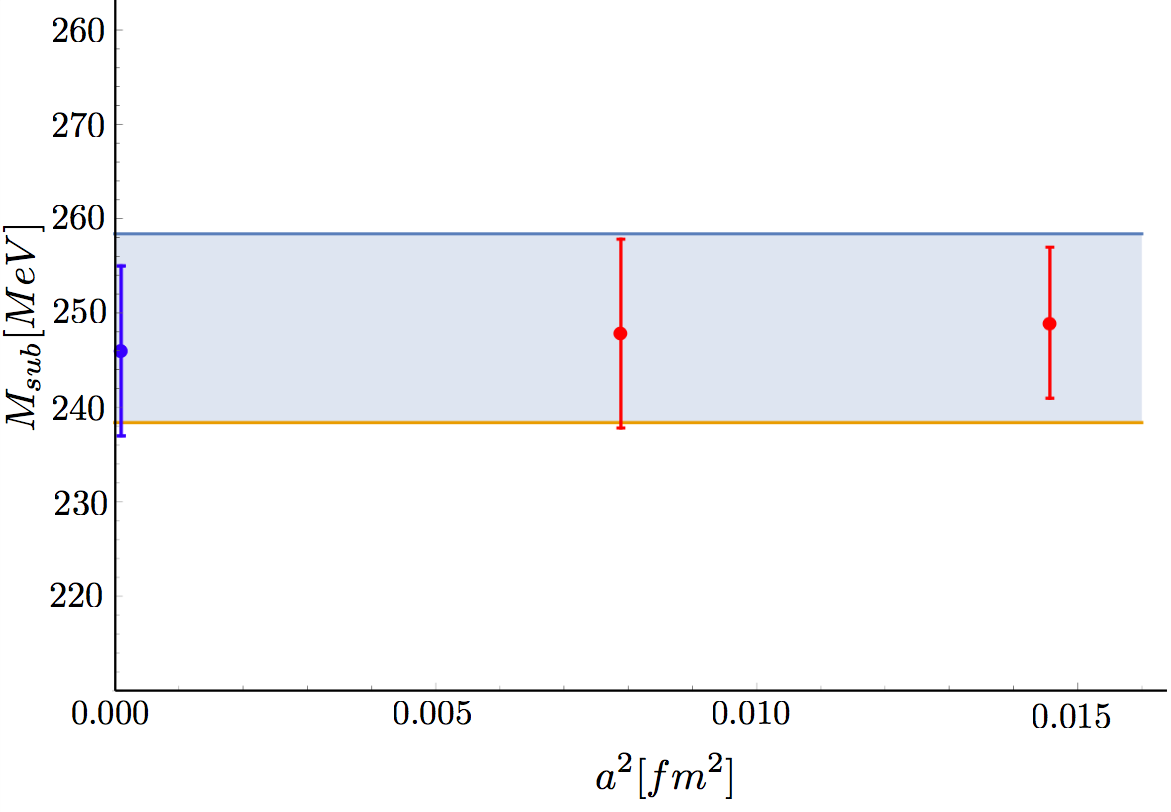}}
	\subfigure[ $\Omega_{cbb}$]{\includegraphics[width=0.45\textwidth,clip]{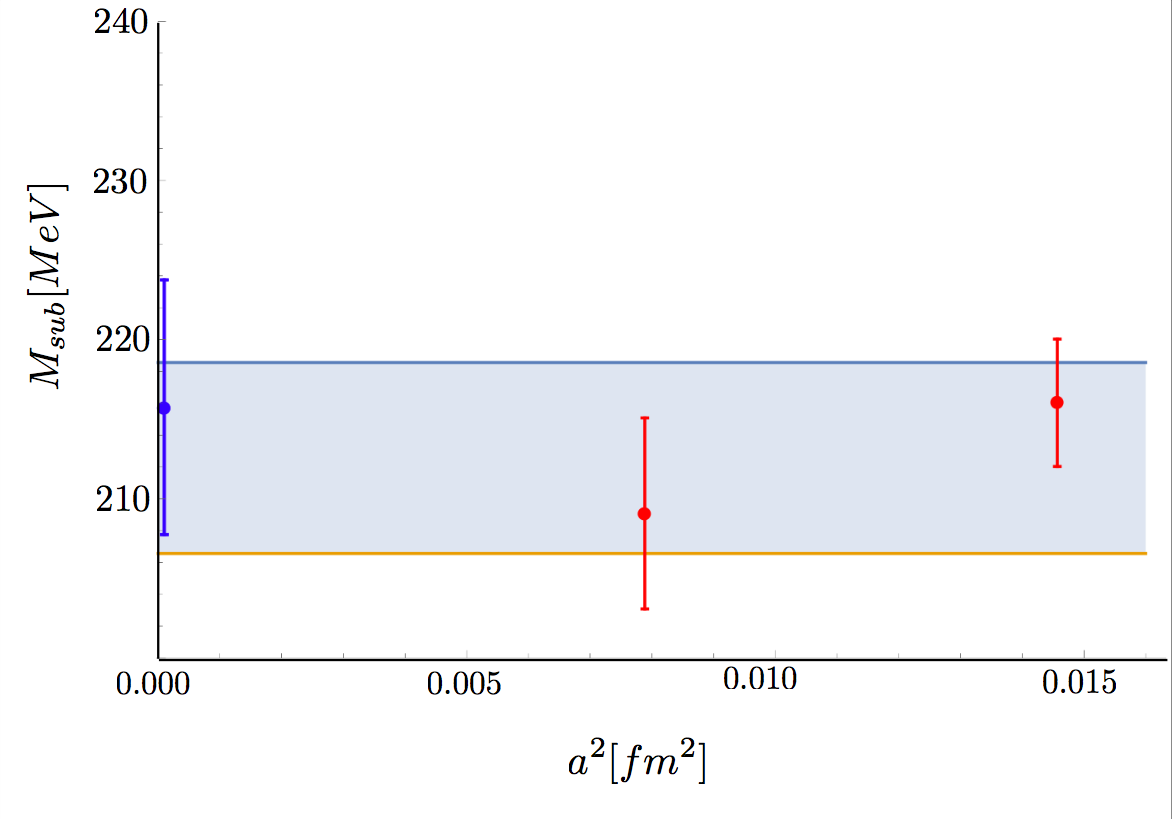}}\hfill
	\subfigure[ $\Omega_{cbb}^{*}$]{\includegraphics[width=0.45\textwidth,clip]{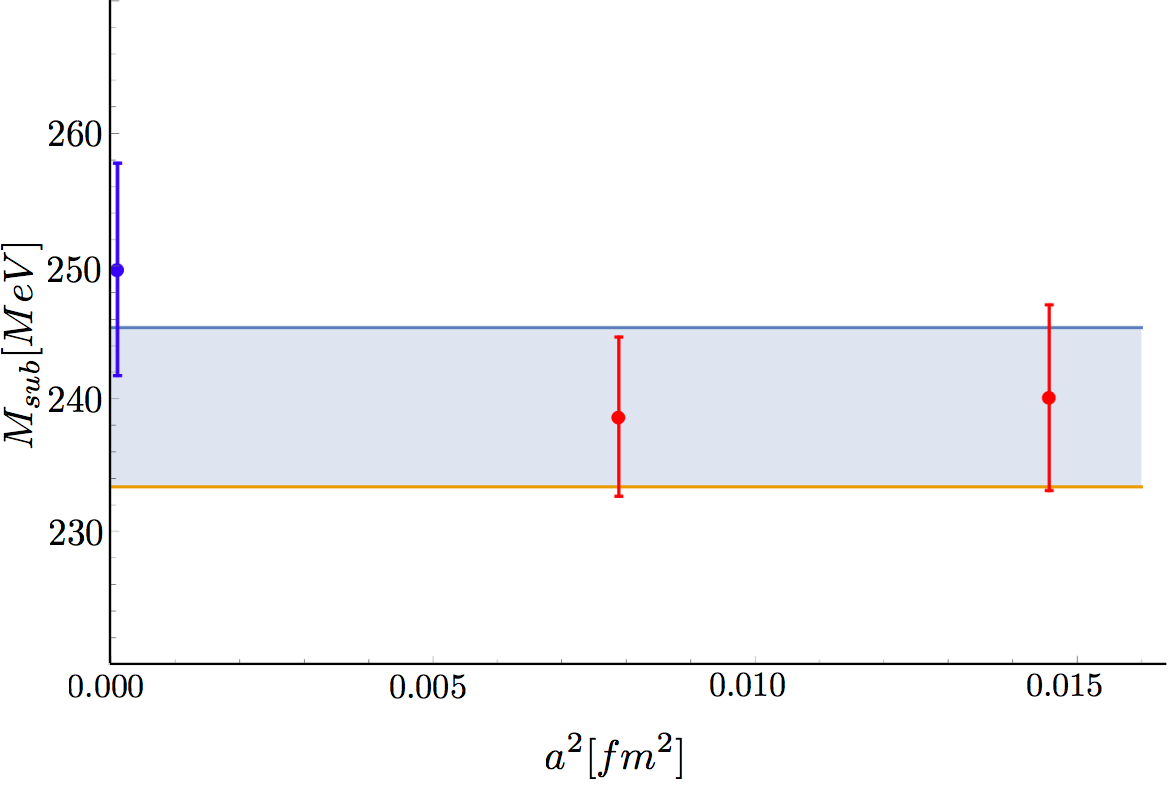}}
	\caption{Subtracted masses for the charmed bottom $\Omega$ baryons. The red points are the preliminary results from this work
	and the blue points are the continuum limit result of Ref.~\cite{Brown:2014ena}}.
	\label{charmedbottomomg}
\end{figure}

To reduce the relative discretization errors due to heavy charm and bottom quarks we present the mass of the $bc$ baryons as : $M_{sub}a = Ma - n_c\overline{Ma}(c\overline{c}) - n_b\overline{Ma}(b\overline{b})$, where $\overline{Ma}(c\overline{c})$ and $\overline{Ma}(b\overline{b})$ are spin-average masses of the 1S charmonia and bottomonia respectively, whereas  $n_c$ and $n_b$ are the number of charm and bottom quarks in $bc$ hadrons.  
 One would expect that these subtractions will effectively remove the heavy quark content and will reduce the discretization errors. In Figure~\ref{charmedbottomomg} we show subtracted energy levels for the charmed-bottom $\Omega$ baryons at two lattice spacings. Our results are compared with those obtained in Ref.~\cite{Brown:2014ena}. Horizontal bars are possible errors one may expect after continuum extrapolation of hadron masses with non-relativistic bottom quarks. In order to obtain physical values we need to add back physical value of $n_c\overline{M}(c\overline{c}) + n_b\overline{M}(b\overline{b})$ to above subtracted masses. In future we will add numbers from another lattice spacing and then will do continuum extrapolation to obtain final numbers for these baryons. 

\vspace*{-0.1in}

\section{Conclusions}\label{sec:details}
We report preliminary results on the ground state energy spectra
for various hadrons with light to bottom quark content. 
We incorporate a unified
approach to treat light to charm quarks uniformly using overlap fermions. A relativistic overlap action is
used on the background of 2+1+1 HISQ
configurations corresponding to lattice spacings of about 0.1207, 0.088
and 0.06 fm. Overlap quark propagators are generated using a multimass
algorithm over a wide range of quark masses corresponding to pseudoscalar mass
from physical pion to about 5.5 GeV. In future we will increase this limit
towards $\eta_b$ using a hyperfine lattice to treat light to bottom quarks with the same action. 
In this calculation for bottom quark
we use a non-relativistic action with non-perturbatively tuned coefficients with terms
up to $\mathcal{O}(v^4)$. For the coarser two ensembles, we
use the values of the improvement coefficients, $c_1$ to $c_6$, as estimated non-perturbatively by the
HPQCD collaboration ~\cite{Dowdall:2011wh}, while for the fine ensemble, we use tree level coefficients.
The charm and bottom masses are tuned by equating the spin-averaged kinetic masses of the $1S$
charmonia and bottomonia states, respectively, to their physical values. 

We present the hyperfine splittings between vector ($1^{-}$) to pseudoscalar ($0^{-}$) mesons as well 
as between $\Delta$-baryon (${3\over 2}^{+}$) to nucleon (${1\over 2}^{+}$) in a large range of quark masses. 
In future we will fit these splittings with appropriate formulae to find out their variations at different quark 
mass ranges. As is well known that these hyperfine splittings provide very useful information about the
spin-spin interactions within the strongly interacting theory and are invaluable 
ingredients for any potential model calculations.  We also show our preliminary results on octet baryons. In future 
we will perform systematic chiral and continuum extrapolations to get 
physical results for these observables. 

We also report our preliminary results on charmed hadrons. Controlling the discretization error 
is a major challenge for heavy quarks and using results at three lattice spacings we are 
able to perform a systematic continuum extrapolation. Extracted value ($115 \pm 3$ MeV) 
of the hyperfine splitting of the $1S$ charmonia agrees very well to its physical value ($113.5 \pm 0.5$ MeV), signifying 
that the discretization errors for charm quark in this calculation is under control. Various mesons 
in the charm-strange sector and as well as in charmonia approach to their physical values correctly after continuum extrapolation. 
We also show our continuum extrapolated results for the ground state of the positive parity singly and doubly charmed 
baryons. We compare our results 
with other lattice results and also with the experimental results where they are available. There is an overall agreement 
with various lattice results for singly charmed baryons and their experimental values.  For the doubly charmed baryons, lattice results, including results from this work, predate the discovery of 
$\Xi_{cc}^{++}$ baryon. These precise and successful predictions of $\Xi_{cc}^{++}$ baryon demonstrate the capability of lattice investigations to rightly guide the future experimental discoveries.  
We also report our preliminary results on hadrons containing both charm and bottom quarks. The hyperfine splitting of 
$B_c$ meson is found to be $55\pm 4$ MeV. For the $bc$ baryons we present results for $\Omega_{ccb}$, $\Omega^*_{ccb}$, 
$\Omega_{cbb}$, and $\Omega^*_{cbb}$. In future we will address other $bc$ baryons including negative parity baryons.

\section{Acknowledgement}
Computations are carried out using computing resources of the Indian Lattice
Gauge Theory Initiative and the Department 
of Theoretical Physics, TIFR. 
We thank A. Salve, K. Ghadiali and P. Kulkarni for technical supports. 
S. M. and N. M. would like to acknowledge support from the Department of Theoretical Physics, TIFR. M. P. acknowledges support from Deutsche Forschungsgemeinschaft Grant No. SFB/TRR 55 and EU under grant no. MSCA-IF-EF-ST-744659 (XQCDBaryons). 
We are grateful to the MILC collaboration and in particular 
to S. Gottlieb for providing us with the HISQ lattices.

\bibliography{Lattice2017_228_Mathur}

\begin{thebibliography}{22}

\bibitem{Aaij:2017ueg}
R.~Aaij et~al. (LHCb), Phys. Rev. Lett. \textbf{119}, 112001 (2017),
  \texttt{1707.01621}

\bibitem{Aaij:2017nav}
R.~Aaij et~al. (LHCb), Phys. Rev. Lett. \textbf{118}, 182001 (2017),
  \texttt{1703.04639}

\bibitem{Bazavov:2012xda}
A.~Bazavov et~al. (MILC), Phys. Rev. \textbf{D87}, 054505 (2013),
  \texttt{1212.4768}

\bibitem{Basak:2012py}
S.~Basak, S.~Datta, M.~Padmanath, P.~Majumdar, N.~Mathur, PoS
  \textbf{LATTICE2012}, 141 (2012), \texttt{1211.6277}

\bibitem{Basak:2013oya}
S.~Basak, S.~Datta, A.T. Lytle, M.~Padmanath, P.~Majumdar, N.~Mathur, PoS
  \textbf{LATTICE2013}, 243 (2014), \texttt{1312.3050}

\bibitem{Lepage:1992nrqcd}
G.P. Lepage~{\em et al.}, Phys.Rev. \textbf{D46}, 4052 (1992), {}

\bibitem{Dowdall:2011wh}
R.J. Dowdall et~al. (HPQCD), Phys. Rev. \textbf{D85}, 054509 (2012),
  \texttt{1110.6887}

\bibitem{Mathur:2016hsm}
N.~Mathur, M.~Padmanath, R.~Lewis, PoS \textbf{LATTICE2016}, 100 (2016),
  \texttt{1611.04085}

\bibitem{Lewis:2008fu}
R.~Lewis, R.M. Woloshyn, Phys. Rev. \textbf{D79}, 014502 (2009),
  \texttt{0806.4783}

\bibitem{Patrignani:2016xqp}
C.~Patrignani et~al. (Particle Data Group), Chin. Phys. \textbf{C40}, 100001
  (2016)

\bibitem{Briceno:2012wt}
R.A. Briceno, H.W. Lin, D.R. Bolton, Phys. Rev. \textbf{D86}, 094504 (2012),
  \texttt{1207.3536}

\bibitem{Alexandrou:2017xwd}
C.~Alexandrou, C.~Kallidonis, Phys. Rev. \textbf{D96}, 034511 (2017),
  \texttt{1704.02647}

\bibitem{Brown:2014ena}
Z.S. Brown, W.~Detmold, S.~Meinel, K.~Orginos, Phys. Rev. \textbf{D90}, 094507
  (2014), \texttt{1409.0497}

\bibitem{Bali:2015lka}
P.~Perez-Rubio, S.~Collins, G.S. Bali, Phys. Rev. \textbf{D92}, 034504 (2015),
  \texttt{1503.08440}

\bibitem{Namekawa:2013vu}
Y.~Namekawa et~al. (PACS-CS), Phys. Rev. \textbf{D87}, 094512 (2013),
  \texttt{1301.4743}

\bibitem{Padmanath:2013bla}
M.~Padmanath, R.G. Edwards, N.~Mathur, M.~Peardon, Proceedings \textbf{Charm}
  (2013), \texttt{1311.4806}

\bibitem{Padmanath:2014bxa}
P.~Madanagopalan, R.G. Edwards, N.~Mathur, M.J. Peardon, PoS
  \textbf{LATTICE2014}, 084 (2015), \texttt{1410.8791}

\bibitem{Padmanath:2015bra}
M.~Padmanath, N.~Mathur, Proceedings \textbf{Charm} (2015), \texttt{1508.07168}

\bibitem{Padmanath:2017lng}
M.~Padmanath, N.~Mathur, Phys. Rev. Lett. \textbf{119}, 042001 (2017),
  \texttt{1704.00259}

\bibitem{Padmanath:2015jea}
M.~Padmanath, R.G. Edwards, N.~Mathur, M.~Peardon, Phys. Rev. \textbf{D91},
  094502 (2015), \texttt{1502.01845}

\bibitem{Mathur:2002ce}
N.~Mathur, R.~Lewis, R.M. Woloshyn, Phys. Rev. \textbf{D66}, 014502 (2002),
  \texttt{hep-ph/0203253}

\bibitem{Lewis:2001iz}
R.~Lewis, N.~Mathur, R.M. Woloshyn, Phys. Rev. \textbf{D64}, 094509 (2001),
  \texttt{hep-ph/0107037}

\end{thebibliography}

\end{document}